\documentstyle[aps,prb]{revtex}
\begin{document}
\draft
\title{One-loop approximation for the Heisenberg antiferromagnet}
\author{A.~Sherman}
\address{Institute of Physics, University of Tartu, Riia
142, 51014 Tartu, Estonia}
\author{M.~Schreiber}
\address{Institut f\"ur Physik, Technische Universit\"at,
D-09107 Chemnitz, Federal Republic of Germany}
\date{\today}
\maketitle
\begin{abstract}
We use the diagram technique for spin operators to calculate Green's
functions and observables of the spin-$\case{1}{2}$ quantum Heisenberg
antiferromagnet on a square lattice. The first corrections to the
self-energy and interaction are taken into account in the chain
diagrams. The approximation reproduces main results of Takahashi's
modified spin-wave theory [Phys.\ Rev.\ B {\bf 40}, 2494 (1989)] and is
applicable in a wider temperature range. The energy per spin calculated
in this approximation is in good agreement with the Monte Carlo and
small-cluster exact-diagonalization calculations in the range $0\leq
T\lesssim 1.2J$ where $J$ is the exchange constant. For the static
uniform susceptibility the agreement is good for $T\lesssim 0.6J$ and
becomes somewhat worse for higher temperatures. Nevertheless the
approximation is able to reproduce the maximum in the temperature
dependence of the susceptibility near $T=0.9J$.
\end{abstract}
\pacs{PACS numbers: 75.30.Ds, 75.50.Ee}

\narrowtext
\section{Introduction}
Properties of the spin-$\case{1}{2}$ quantum Heisenberg antiferromagnet
on a square lattice have attracted much attention in connection with the
investigation of cuprate-perovskite high-temperature superconductors.
Much of this interest stems from Anderson's original suggestion
\cite{anderson87} that quantum spin fluctuation in CuO$_2$ planes of
these compounds may be responsible for superconductivity. At present it
is generally accepted that the undoped CuO$_2$ planes are well described
by the Heisenberg model with nearest-neighbor interaction (see
Ref.~\onlinecite{birg89} and references therein). 

Numerical calculations \cite{reger,liang} and the analysis of
experimental data \cite{chakra} presented strong evidence that the
two-dimensional nearest-neighbor $s=\case{1}{2}$ Heisenberg
antiferromagnet has long-range order at zero temperature. For $T\neq 0$
the Hohenberg-Mermin-Wagner theorem \cite{mermin} shows rigorously that
long-range order is destroyed. In real quasi-two-dimensional cuprate
perovskites long-range order persists for temperatures lower than the
N\'eel temperature. In this case the ordering is destroyed by small
concentrations of carriers $x\lesssim 0.01$. \cite{birg89} In both cases
there is a range of parameters where the arising short-range order is
characterized by the magnetic correlation length $\xi$ which is much
larger than the in-plane intersite distance $a$. In this situation, the
spin-wave theories \cite{takahashi,tang,sherman} modified for
short-range order may be a good starting point for the investigation of
antiferromagnets. \cite{schreiber} For undoped crystals observables
calculated in these theories agree nicely (cp.\
Refs.~\onlinecite{takahashi,tang,sherman}) with the small-cluster
exact-diagonalization and Monte Carlo data up to temperature $T \approx
0.6J$ where $J$ is the exchange constant. This corresponds to the
correlation length $\xi \approx 10a$. \cite{takahashi,tang} However, in
cuprates for the region of carrier concentrations $x\gtrsim 0.1$, which
is of prime interest, $\xi$ is of the order of few lattice spacings.
\cite{zha} In this case the application of the modified spin-wave
theories (MSWT) becomes doubtful.

In this paper we extend the analytic description of elementary
excitations of the undoped antiferromagnet to the region of higher
temperatures and shorter correlation lengths. For this purpose we use
the diagram technique for spin operators, developed in
Ref.~\onlinecite{vaks} (see also Ref.~\onlinecite{izyumov}). We consider
the first corrections to the simplest chain diagrams. Due to peculiar
shapes of the main correction diagrams this approximation will be
referred to as the one-loop approximation (OLA). We note that the OLA is
not rotationally invariant (neither is the MSWT).

As will be seen below, the obtained formulas are in many respects
similar to the formulas of the MSWT developed in
Refs.~\onlinecite{takahashi,tang}. The major difference between these
two approximations is in the evaluation of the excitation frequency. In
the MSWT two parameters defining the frequency are deduced from the
constraint of zero site magnetization and a self-consistency condition,
while in the OLA these parameters are determined from the values of
diagrams. The OLA reproduces the main results of the MSWT and is
applicable in a wider temperature range. The energy per spin calculated
in the OLA agrees nicely with the Monte Carlo and small-cluster
exact-diagonalization calculations up to $T\approx 1.2J$. In the
temperature range $T\lesssim 0.6J$ both approximations give values of
the static uniform susceptibility in good agreement with numerical
calculations. In the OLA, the agreement becomes somewhat worse for
larger temperatures. Nevertheless this approximation is able to
reproduce the maximum in the temperature dependence of the
susceptibility near $T=0.9J$ in close correspondence with experiment and
numerical calculations, while in the MSWT the susceptibility grows
monotonously with temperature. 

It is worth noting also that the OLA reproduces results of the
traditional spin-wave approximation on the zero-temperature
renormalization of the excitation frequency and of the renormalization
group theory \cite{chakra} on the temperature variation of the
correlation length. We have found also good agreement between data on
zero-temperature spin correlations obtained with the projected Monte
Carlo simulations, the MSWT of Ref.~\onlinecite{sherman}, and the
present OLA.

The outline of the paper is as follows. In Sec.~II the diagram technique
for spin operators is discussed. The series of the simplest chain
diagrams is summed in Sec.~III\@. At $T=0$ this approximation is similar
to the spin-wave approximation suggested by Anderson. \cite{anderson}
The one-loop corrections to the chain diagrams are considered in
Sec.~IV\@. The obtained excitation spectrum is compared here and in
Sec.~V with the spectra of the traditional spin-wave theory and the
MSWT\@. The spin-wave approximation based on the obtained formulas is
discussed in Sec.~V\@. Here results of the OLA for the energy,
susceptibility and spin correlations are compared with
exact-diagonalization, Monte Carlo, and MSWT data. Our summary is given
in Sec.~VI.

\section{Diagram technique}
The Hamiltonian of the Heisenberg antiferromagnet can be written in the
form
\begin{eqnarray}
H&=&\sum_{\bf lm}J_{\bf lm}{\bf s_ls_m}\nonumber\\
 &=&\sum_{\bf lm}J_{\bf lm}\left[s^z_{\bf l}s^z_{\bf m}
 +\frac{1}{2}\left(s^+_{\bf l}s^-_{\bf m}+s^-_{\bf l}s^+_{\bf m}
 \right)\right],\label{hamilton}
\end{eqnarray}
where {\bf l} and {\bf m} label the sites of the two sublattices of the
square lattice, the exchange constants $J_{\bf lm}$ are expected to be
nonzero for nearest neighbor sites only, $J_{\bf lm}=J\sum_{\bf
a}\delta_{\bf l,m+a}$ with ${\bf a}=(0,\pm a), (\pm a,0)$, ${\bf s_l}$
is the spin-$\case{1}{2}$ operator the components of which satisfy the
commutation relations
\begin{equation}
\left[s^x_{\bf n},s^y_{\bf n'}\right]=is^z_{\bf n}\delta_{\bf nn'},
\label{commutation}\end{equation}
and analogously for cyclic permutations of indices $x$, $y$ and $z$.
Here ${\bf n}=\bf l$ or $\bf m$ and $s^{\pm}_{\bf n}=s^x_{\bf n}\pm 
is^y_{\bf n}$. For $s=\case{1}{2}$ we have $s^z_{\bf n}=-\case{1}{2}+
s^+_{\bf n}s^-_{\bf n}$ and $s^+_{\bf n}s^-_{\bf n}+s^-_{\bf n}
s^+_{\bf n}=1$.

On one sublattice we change notations: $s^{\mp}_{\bf
m}=-\tilde{s}^{\pm}_{\bf m}$ and $s^z_{\bf m}=-\tilde{s}^z_{\bf m}=
\case{1}{2}-\tilde{s}^+_{\bf m}\tilde{s}^-_{\bf m}$. The new operators
$\tilde{s}^\alpha_{\bf m}$ satisfy commutation relations
(\ref{commutation}). Substituting the new notations in Hamiltonian
(\ref{hamilton}) and omitting the tildes we find
\begin{eqnarray}
&&H=-\frac{JN}{2}+H_0+H_1, \nonumber\\
&&H_0=2J\left(\sum_{\bf l}s^+_{\bf l}s^-_{\bf l}+
 \sum_{\bf m}s^+_{\bf m}s^-_{\bf m}\right), \label{newh}\\
&&H_1=-\sum_{\bf lm}J_{\bf lm}\left[s^+_{\bf l}s^-_{\bf l}
 s^+_{\bf m}s^-_{\bf m}+\frac{1}{2}\left(s^+_{\bf l}s^+_{\bf m}+
 s^-_{\bf l}s^-_{\bf m}\right)\right], \nonumber
\end{eqnarray}
where $N$ is the number of lattice sites.

Our aim is the calculation of Green's functions
\begin{eqnarray*}
&&D_{\bf ll'}(\tau)=-\langle{\cal P}s^-_{\bf l}(\tau)s^+_{\bf l'}
\rangle,\\
&&D'_{\bf lm}(\tau)=-\langle{\cal P}s^-_{\bf l}(\tau)s^-_{\bf m}
\rangle,\\
&&D''_{\bf ml}(\tau)=-\langle{\cal P}s^+_{\bf m}(\tau)s^+_{\bf l}
\rangle,
\end{eqnarray*}
where $s^-_{\bf l}(\tau)=\exp(H\tau)s^-_{\bf l}\exp(-H\tau)$, the
time-ordering operator ${\cal P}$ arranges operators in order of
increasing time from right to left, and angular brackets denote
thermodynamic averaging with the Hamiltonian $H$. It follows from the
above definition that $\left[D'_{\bf lm}(\tau)\right]^*=D''_{\bf
ml}(\tau)$.

Using $H_0$ and $H_1$ from Eq.~(\ref{newh}) as the unperturbed
Hamiltonian and perturbation, respectively, Green's fun\-c\-ti\-ons can
be represented by an infinite series generated by the known series of
the evolution operator. The terms of the former series contain Green's
functions
\begin{equation}
\langle{\cal P}s^{\alpha_1}_{{\bf n}_1}(\tau_1)
s^{\alpha_2}_{{\bf n}_2}(\tau_2)\ldots s^{\alpha_p}_{{\bf
n}_p}(\tau_p)\rangle_0
\label{term}\end{equation}
where the averaging and time dependences of the operators are determined
with the unperturbed Hamiltonian $H_0$ as indicated by the subscript 0
of the angular brackets, so that
$$s^\pm_{\bf n}(\tau)=s^\pm_{\bf n}e^{\pm 2J\tau}.$$
As follows from Eq.~(\ref{newh}), Green's functions (\ref{term}) contain
only operators $s^+$ and $s^-$ (i.e.\ $\alpha_i=+$ or $-$) and do not
contain operators $s^z$. These Green's functions are nonzero when the
number of $s^-$ operators is equal to the number of $s^+$ operators. To
calculate these functions we use Wick's theorem proposed for spin
operators in Ref.~\onlinecite{vaks} (see also
Ref.~\onlinecite{izyumov}). The theorem can be written in the form
\begin{eqnarray}
&&\langle{\cal P}s^{\alpha_1}_{{\bf n}_1}(\tau_1)\ldots 
 s^-_{\bf n}(\tau)\ldots s^{\alpha_p}_{{\bf n}_p}(\tau_p)\rangle_0
 \nonumber\\
&&\mbox{ }=K(\tau-\tau_1)\langle{\cal P}\left[s^-_{\bf 
 n},s^{\alpha_1}_{{\bf n}_1}\right]_{\tau_1}\ldots s^{\alpha_p}_{{\bf 
 n}_p}(\tau_p)\rangle_0+\ldots\nonumber\\
&&\mbox{ }+K(\tau-\tau_p)\langle{\cal P}
 s^{\alpha_1}_{{\bf n}_1}(\tau_1)\ldots \left[s^-_{\bf n},
 s^{\alpha_p}_{{\bf n}_p}\right]_{\tau_p}\rangle_0 \label{wick}
\end{eqnarray}
where
\[ K(\tau)=e^{-2J\tau}
 \left(1-e^{-2J\beta}\right)^{-1}\left\{\begin{array}{ll}
 1, & \tau>0 \\
 e^{-2J\beta}, & \tau<0
 \end{array}\right. \]
and $\beta^{-1}=T$ is the temperature. The subscripts $\tau_i$ of the
commutators in Eq.~(\ref{wick}) are the time arguments of the operators
arising after the commutations. Equation~(\ref{wick}) is easily verified
by transferring $s^-_{\bf n}(\tau)$ to the right averaging bracket with
the introduction of the commutators, using cyclic permutation in the
averaging and transferring the operator from the left bracket to its
initial position.

As noted above, terms (\ref{term}) in the series expansion of Green's
functions contain only operators $s^-$ and $s^+$. However, after the
application of Wick's theorem~(\ref{wick}) operators $s^z$ appear in the
averaging. Equation~(\ref{wick}) is also applicable, if some of the
$s^{\alpha_i}$ are $s^z$ operators and the couplings of $s^-$ with $s^z$
have to be taken into account along with the coupling of the former
operators with $s^+$. Equation~(\ref{wick}) is used until only operators
$s^z$ are left in the averaging brackets. Since the Hamiltonian $H_0$ is
the sum of terms related to individual sites, these averages are split
into averages related to individual sites. Such averages are easily
calculated using the recurrence relation
\begin{eqnarray}
&&\left\langle\left(s^z_{\bf n}\right)^{k+1}\right\rangle_0=
 \left\langle\left(s^z_{\bf n}\right)^k\right\rangle_0
 \left\langle s^z_{\bf n}\right\rangle_0
 -\frac{1}{2\beta}\frac{\partial}{\partial J}
 \left\langle\left(s^z_{\bf n}\right)^k\right\rangle_0,\nonumber\\
&&\label{recurrence}\\
&&\langle s^z_{\bf n}\rangle_0=-\frac{1}{2}\tanh\left(J\beta\right)
 \equiv-\frac{c}{2}.\nonumber
\end{eqnarray}
For the following discussion we notice that
\begin{mathletters}
\begin{eqnarray}
&&\left\langle s^z_{\bf n}s^z_{\bf n'}\right\rangle_0=
 \left\langle s^z_{\bf n}\right\rangle_0^2-
 \frac{1}{2\beta}\frac{\partial}{\partial J}
 \left\langle s^z_{\bf n}\right\rangle_0\delta_{\bf nn'}, \\
&&\left\langle s^z_{\bf n}s^z_{\bf n'}s^z_{\bf n''}\right\rangle_0=
 \left\langle s^z_{\bf n}\right\rangle_0^3 \nonumber\\
&&\quad-\left\langle s^z_{\bf n}\right\rangle_0
 \frac{1}{2\beta}\frac{\partial}{\partial J}
 \left\langle s^z_{\bf n}\right\rangle_0\left(\delta_{\bf nn'}
 +\delta_{\bf nn''}+\delta_{\bf n'n''}\right)\nonumber\\
&&\quad+\frac{1}{4\beta^2}\frac{\partial^2}{\partial J^2}
 \left\langle s^z_{\bf n}\right\rangle_0\delta_{\bf nn'}
 \delta_{\bf nn''}, 
\end{eqnarray}
\label{szs}\end{mathletters}
where we took into account that $\left\langle s^z_{\bf
n}\right\rangle_0$ and its derivatives do not depend on {\bf n}.
Analogous formulas can be written for an average of a larger number of
$s^z$ operators.

The diagram technique can be used to visualize the terms of the
perturbation theory. \cite{vaks,izyumov} $K(\tau-\tau')$ will be
displayed graphically by a solid line with an arrow directed from $\tau$
to $\tau'$ [from the respective $s^-$ to $s^+$ or $s^z$ operators, see
Eq.~(\ref{wick})]. An $s^z$ operator is connected with the terminal
point of this line unless a second line with an arrow enters into this
point. This latter case corresponds to the situation when an $s^z$
operator which appears after the commutation of $s^-$ and $s^+$
operators participates in another coupling. The $s^-$ operator produced
in this way participates in further couplings and therefore a third line
has to issue from the terminal point of the former two lines (see
Fig.~\ref{figi}a). 

Excluding two free points labelled by the time arguments $\tau$ and $0$
all other initial and terminal points of the directed lines are
connected by wavy lines corresponding to the exchange integrals $J_{\bf
lm}$ in the longitudinal [the first term in $H_1$, Eq.~(\ref{newh})] and
transverse interactions (the second and third terms in $H_1$). The
transverse interaction line connects two directed lines, either
terminating or initiating (Fig.~\ref{figi}b and c). The longitudinal
interaction line connects four directed lines, one terminating and one
initiating at each end of the interaction line (Fig.~\ref{figi}d).

The delta functions in the last terms of Eqs.~(\ref{szs}a)
and~(\ref{szs}b) introduce limitations on the site indices of the $s^z$
operators. In the diagrams we shall denote these limitations by dashed
lines connecting ends of directed lines (points corresponding to $s^z$
operators). A second-order diagram with such a dashed line arising due
to the second term on the right-hand side of Eq.~(\ref{szs}a) is shown
in Fig.~\ref{figi}e. As usual the partition function in the denominators
of Green's functions cancels all unlinked diagrams. It must be noted
that a dashed line can connect an otherwise unlinked diagram, as shown
in Fig.~\ref{figi}e.

\section{Chain diagrams}
A satisfactory description of low-temperature excitations of the
Heisenberg antiferromagnet is achieved in the Anderson approximation
\cite{anderson} which neglects the longitudinal interaction and
substitutes the spin operators $s^+$ and $s^-$ by Boson annihilation and
creation operators. In the diagram representation this approximation is
reduced to neglecting diagrams with longitudinal interaction lines and
with dashed lines or loops formed by directed and transversal wavy lines
(these two latter elements arise due to the non-Boson statistics of spin
operators; it has to be noted that the loops contain the ``triple
points'' similar to that shown in Fig.~\ref{figi}a). The remaining chain
diagrams for $D_{\bf ll'}(\tau)$ are shown in Fig.~\ref{figii}.

After the Fourier transformation
$$D({\bf k},i\omega_n)=\int_0^\beta\!\! d\tau\sum_{\bf l}e^{i\omega_n\tau}
 e^{i{\bf k}({\bf l}-{\bf l'})}D_{\bf ll'}(\tau),$$
where $\omega_n=\pi Tn$ is Matsubara's frequency with an even integer
$n$, the series of chain diagrams can be written as
\begin{eqnarray}
D^c({\bf k},i\omega_n)&=&-cK(i\omega_n)\Biggl\{1+\left(
 \frac{cJ_{\bf k}}{2}\right)^2K(-i\omega_n)K(i\omega_n)\nonumber\\
&&+\left[\left(\frac{cJ_{\bf k}}{2}\right)^2K(-i\omega_n)
 K(i\omega_n)\right]^2+\ldots\Biggr\}\nonumber\\
&=&\frac{c\left(i\omega_n+2J\right)}{\left(i\omega_n\right)^2-
 \omega^2_{\bf k}},\nonumber\\
D'^c({\bf k},i\omega_n)&=&-\frac{J_{\bf k}}{2}c^2K(i\omega_n)
 K(-i\omega_n) \label{chain}\\
&&\times\Biggl\{1+\left(\frac{cJ_{\bf k}}{2}\right)^2
 K(-i\omega_n)K(i\omega_n)\nonumber\\
&&+\left[\left(\frac{cJ_{\bf k}}{2}\right)^2K(-i\omega_n)
 K(i\omega_n)\right]^2+\ldots\Biggr\}\nonumber\\
&=&\frac{1}{2}\frac{c^2J_{\bf k}}{\left(i\omega_n\right)^2-
 \omega^2_{\bf k}},\nonumber
\end{eqnarray}
where
$$K(i\omega_n)=-\left(i\omega_n-2J\right)^{-1},\quad
J_{\bf k}=4J\gamma_{\bf k}$$
are the Fourier transforms of $K(\tau)$ and $J_{\bf lm}$, respectively,
$\gamma_{\bf k}=\sum_{\bf a}\exp(i{\bf ka})/4$ and the excitation 
frequency
\begin{equation}
\omega_{\bf k}=2J\sqrt{1-c^2\gamma^2_{\bf k}}.
\label{swf}\end{equation}
At $T=0$ when $c=1$ [see Eq.~(\ref{recurrence})] Eq.~(\ref{swf})
reproduces the
spin-wave spectrum of Ref.~\onlinecite{anderson}. For nonzero
temperature $c<1$ and a gap opens near ${\bf k}=0$. This gap is
exponentially small for low temperatures $J\beta\gg 1$.

\section{First corrections}
The first corrections to the chain diagrams contain one longitudinal
interaction line, a loop with the ``triple point'' or a dashed line in
the self-energy or in the renormalized interaction. Similar corrections
were considered for the case of a three-dimensional antiferromagnet and
zero temperature in Ref.~\onlinecite{wang} with the use of another
diagram technique and for a ferromagnet in
Refs.~\onlinecite{vaks,izyumov,lewis}. The respective diagrams are shown
in Fig.~\ref{figiii}. Diagrams (a) and (b) are self-energy corrections
which are inserted in directed lines of the chain diagrams. Following
Ref.~\onlinecite{wang} we carried out partial summations in these
diagrams: the line with two arrows indicates the series of chain
diagrams (\ref{chain}). Diagrams (d) to (f) are related to terminal
points of directed lines and we include them together with diagram (c)
to the renormalization of the interaction lines in the chain diagrams.
However, in the chain diagrams the number of directed lines exceeds by
one the number of wavy lines. Therefore the renormalized Green's
functions have the additional multiplier $\alpha=1+\sigma_d+\sigma_e+
\sigma_f$ where $\sigma_i$ is the contribution of the $i$th diagram in
Fig.~\ref{figiii}. Notice that the multiplier $\alpha$ was not included
in Refs.~\onlinecite{wang,lewis}.

With these corrections Green's functions read
\begin{eqnarray}
&&D({\bf k},i\omega_n)=\frac{\alpha c\left(i\omega_n+2J\varepsilon
 \right)}{\left(i\omega_n\right)^2-\Omega^2_{\bf k}},\nonumber\\[-1ex]
&&\label{gf}\\[-1ex]
&&D'({\bf k},i\omega_n)=\frac{2J\alpha c^2\phi\gamma_{\bf k}}
 {\left(i\omega_n\right)^2-\Omega^2_{\bf k}},\nonumber
\end{eqnarray} 
where
\begin{eqnarray} 
&&\Omega_{\bf k}=2J\sqrt{\varepsilon^2-\left(c\phi\gamma_{\bf k}
 \right)^2},\label{freq}\\
&&\varepsilon=1+c^2(1-I_1)+cI,\label{eps}\\
&&\phi=1+c^2I+\frac{1}{c^2}(1-cI_1)\nonumber\\
&&\quad-\cosh^{-2}(J\beta)I'+J\beta\cosh^{-2}(J\beta)I,
 \label{phi}\\
&&\alpha=1+\frac{1}{c^2}(1-cI_1)-\cosh^{-2}(J\beta)I'\nonumber\\
&&\quad+J\beta\cosh^{-2}(J\beta)I, \label{alpha} 
\end{eqnarray}
\begin{mathletters}
\begin{eqnarray}
&&I=\frac{1}{2J}\frac{2}{N}\sum_{\bf k}\left(\frac{J_{\bf 
 k}}{2}\right)^{\!\! 2}\!\frac{1}{\omega_{\bf k}}\coth\left(
 \frac{\beta\omega_{\bf k}}{2}\right),\\
&&I_1=2J\frac{2}{N}\sum_{\bf k}\frac{1}{\omega_{\bf k}}
 \coth\left(\frac{\beta\omega_{\bf k}}{2}\right),\\
&&I'=\frac{2}{N}\sum_{\bf k}\left(\frac{J_{\bf k}}{2}\right)^{\!\! 2}
 \!\frac{1}{(i\omega_n)^2-\omega_{\bf k}^2}.
\end{eqnarray}
\label{sums}\end{mathletters}
Here and below summations over wave vectors are carried over the
magnetic Brillouin zone. The second and third terms on the right-hand
side of Eq.~(\ref{eps}) are contributions of diagrams (a) and (b) in
Fig.~\ref{figiii}, respectively. The second to fifth terms on the
right-hand side of Eq.~(\ref{phi}) are contributions of diagrams (c) to
(f), respectively. As mentioned, diagrams (d), (e), and (f) give the
last three terms in $\alpha$, Eq.~(\ref{alpha}). In the calculation of
the correction introduced by diagram (c) we took into account that for
any $f({\bf k})$ invariant with respect to operations of the crystal
point group
$$\sum_{\bf k'}J_{\bf k-k'}f({\bf k'})=J_{\bf k}J^{-1}_{\bf 0}
 \sum_{\bf k'}J_{\bf k'}f({\bf k'}).$$

To estimate the contributions of the diagrams in Fig.~\ref{figiii} we
introduce the function \cite{takahashi}
\begin{mathletters}
\begin{equation}
w(x)=\frac{2}{N}\sum_{\bf k}\delta\left(x-\gamma_{\bf k}\right),\;
 0\leq x\leq 1,
\end{equation}
and consider an infinite square lattice. In this case
\begin{equation}
w(x)=\left(\frac{2}{\pi}\right)^2{\cal K}\left(\sqrt{1-x^2}\right)=
 \frac{2}{\pi}+{\cal O}(1-x),
\end{equation}
\label{w}\end{mathletters}
where ${\cal K}(k)$ is a complete elliptic integral of the first kind
with modulus $k$. For low temperatures $T\ll J$ we find
\begin{eqnarray}
&&I=c^{-2}(I_1-I_2),\nonumber\\
&&I_1=\int^1_0\frac{w(x)}{\sqrt{1-c^2x^2}}\coth\!\left(\beta J
 \sqrt{1-c^2x^2}\right)dx\nonumber\\
&&\quad=w(1)+2(1-m_0)-\frac{w(1)}{\beta J}\ln(4\beta J)+{\cal O}\left[
 (\beta J)^{-3}\right],\nonumber\\
&&I_2=\int^1_0w(x)\sqrt{1-c^2x^2}\coth\!\left(\beta J\sqrt{1-c^2x^2}
 \right)dx\label{ints}\\
&&\quad=2(1-m_1)+\frac{w(1)\zeta(3)}{2(\beta J)^3}+{\cal O}\left[
 (\beta J)^{-5}\right],\nonumber\\
&&I'=c^{-2}\int^1_0\frac{w(x)x^2}{x^2-x^2_0}dx\nonumber\\
&&\quad=\frac{w(1)x_0}{2}\ln\left|\frac{1-x_0}{1+x_0}\right|+
 {\cal O}\left[(\beta J)^0\right],\nonumber
\end{eqnarray} 
where
\begin{eqnarray}
&&m_0=1-\frac{1}{2}\int^1_0\frac{w(x)}{\sqrt{1-x^2}}dx\approx 0.303398,
 \nonumber\\
&&m_1=1-\frac{1}{2}\int^1_0w(x)\sqrt{1-x^2}dx\approx 0.578974,
 \label{pars}\\
&&x_0=\frac{\sqrt{\omega^2_n+4J^2}}{2Jc},\nonumber
\end{eqnarray}
$\zeta(x)$ is the Riemann zeta function, and ${\cal O}\left[(\beta
J)^0\right]$ in $I'$ comprises terms slowly varying with $\omega_n$.

From Eqs.~(\ref{phi}), (\ref{alpha}), (\ref{ints}), and~(\ref{pars}) one
can see that due to the multipliers $\cosh^{-2}(J\beta)$ for low
temperatures, where the obtained formulas are expected to be valid,
$\sigma_e$ and $\sigma_f$ are much smaller than $\sigma_c$ and
$\sigma_d$ and therefore the two former contributions can be omitted.
The retained diagrams contain a closed loop formed by directed and wavy
lines [diagrams (b), (c), and (d)] or a bubble [diagram (a)]. Due to
these shapes of the main diagrams we use the term ``one-loop
approximation''. From the above equations we find that in two dimensions
contributions of these separate diagrams are not small in comparison
with the unperturbed self-energy and interaction. For example, the
zero-temperature value of diagram (b) is $2.38J$ to be compared with the
unperturbed self-energy $2J$. However, the sums of the values of the
diagrams (a) and (b), as well as (c) and (d) are small in comparison
with the respective unperturbed values. Substituting (\ref{ints}) in
Eqs.~(\ref{freq})--(\ref{alpha}) we find for an infinite lattice
\begin{eqnarray}
&&\Omega_{\bf k}=2J\varepsilon\sqrt{1-r^2\gamma^2_{\bf k}},
 \nonumber\\
&&\varepsilon=2m_1-\frac{w(1)\zeta(3)}{2(\beta J)^3}+{\cal 
 O}\left[(\beta J)^{-5}\right],\nonumber\\
&&r=c\phi\varepsilon^{-1}=1-2m_1^{-1}\bigl\{1+2w(1) \label{omega}\\
&&\quad-4m_0+2m_1+{\cal O}\left[
 (\beta J)^{-1}\right]\bigr\}\exp(-2J\beta),\nonumber\\
&&\alpha=2m_0-w(1)+{\cal O}\left[(\beta J)^{-1}\right].\nonumber
\end{eqnarray}
The zero-temperature renormalization of the frequencies of elementary
excitations, given by Eq.~(\ref{omega}), coincides with the result
obtained previously in the traditional spin-wave approximation
\cite{oguchi,igarashi} and by other methods.
\cite{takahashi,tang,sherman} The temperature variation of the frequency
described by the term $-0.5w(1) \zeta(3)(\beta J)^{-3}$ in
Eq.~(\ref{omega}) is close to the MSWT result \cite{takahashi}
$-0.322w(1)\zeta(3)(\beta J)^{-3}$. For nonzero temperature $r<1$ which
produces a gap near ${\bf k}=0$ in the spectrum of elementary
excitations. For low temperatures the size of the gap is exponentially
small and the obtained exponent $-J\beta$ is close to the MSWT value
\cite{takahashi} $-4m_0m_1w^{-1}(1)J\beta \approx -1.1037J\beta$.
However, the preexponential factors are different in these two
approaches.

\section{The spin-wave approximation}
In contrast to the small correction to the excitation frequency, the
numerators of Green's functions (\ref{gf}) are essentially renormalized
in comparison with the chain approximation (\ref{chain}). In
Eq.~(\ref{omega}) $\alpha \approx -0.03$ and with this small multiplier
Green's functions (\ref{gf}) nearly vanish. It is worth noting the
direct analogy of this result with the MSWT \cite{takahashi,tang} where
the correlations $\left\langle s^-_{\bf l}s^+_{\bf l'}\right\rangle$,
$\left\langle s^-_{\bf l}s^-_{\bf m}\right\rangle$ and the related
Green's functions are exactly zero. Thus, like this theory, the OLA is
not rotationally invariant and the $z$ components of the spin
correlations are much larger than the other components. In the following
discussion these latter components will be neglected.

The analogy with the MSWT may be continued, if attention is drawn to the
fact that the quantities
$$d({\bf k},i\omega_n)=D({\bf k},i\omega_n)/\alpha,\quad
d'({\bf k},i\omega_n)=D'({\bf k},i\omega_n)/\alpha$$
are Green's functions of spin waves. Indeed, the factor $\alpha$ in
Green's functions (\ref{gf}) differs from 1 due to the terminal-point
diagrams (d), (e), and (f) which account for the non-Boson statistics of
spin operators corresponding to the terminal points. Thus, dropping
$\alpha$ in Green's functions (\ref{gf}) corresponds to the replacement
of the spin operators by the respective Boson operators of spin waves.
We introduce these latter operators with the Dyson-Maleev transformation
\begin{eqnarray*}
&&s^-_{\bf l}=b^\dagger_{\bf l},\; s^+_{\bf l}=\left(1-b^\dagger_{\bf l}
 b_{\bf l}\right)b_{\bf l},\; s^z_{\bf l}=\frac{1}{2}-b^\dagger_{\bf l}
 b_{\bf l},\\
&&s^-_{\bf m}=-b_{\bf m},\; s^+_{\bf m}=-b^\dagger_{\bf m}\left(1-
 b^\dagger_{\bf m}b_{\bf m}\right),\; s^z_{\bf m}=-\frac{1}{2}+
 b^\dagger_{\bf m}b_{\bf m}
\end{eqnarray*}
(here and below we return to the initial coordinate system on the
sublattice labelled by the index {\bf m}). In these notations $d$ and
$d'$ acquire the familiar form of the magnon Green's functions $$d_{\bf
ll'}(\tau)=-\left\langle{\cal P}b^\dagger_{\bf l}(\tau) b_{\bf
l'}\right\rangle,\quad d'_{\bf lm}(\tau)=-\left\langle{\cal P}
b^\dagger_{\bf l}(\tau)b^\dagger_{\bf m}\right\rangle.$$ It is essential
to note that the self-energy and interaction-line corrections, which are
also connected with the non-Boson statistics of spin operators and
renormalize the excitation frequency, are taken into account in $d$ and
$d'$. Analogous corrections are allowed for in the MSWT and, as a
consequence, $d$ and $d'$ derived from Eq.~(\ref{gf}) up to the factor
$c$ coincide in their form with the magnon Green's functions obtained
\cite{sherman} in that theory. However, in the discussed spin-wave
approximation the excitation frequency is determined by
Eqs.~(\ref{freq})--(\ref{phi}), and (\ref{sums}). They differ from the
equations \cite{takahashi,tang} defining the frequency in the MSWT.

Taking into account that 
\begin{eqnarray*}
&&\left\langle b^\dagger_{\bf l}b_{\bf l'}\right\rangle=-d_{\bf 
 ll'}(\tau=+0)=-\int_{\cal C}\frac{e^{-z\tau}}{1-e^{-z\beta}}
 d_{\bf ll'}(z)\frac{dz}{2\pi i},\\
&&\left\langle b_{\bf l'}b^\dagger_{\bf l}\right\rangle=-d_{\bf 
 ll'}(\tau=-0)=-\int_{\cal C}\frac{e^{-z\tau}}{e^{z\beta}-1}
 d_{\bf ll'}(z)\frac{dz}{2\pi i},
\end{eqnarray*}
where ${\cal C}$ is a closed contour embracing the complex plane, we
find
\widetext
\begin{eqnarray}
&&\left\langle {\bf s_ls_{l'}}\right\rangle=
 \left\langle s^z_{\bf l}s^z_{\bf l'}\right\rangle\approx
 \left\langle b^\dagger_{\bf l}b_{\bf l'}\right\rangle
 \left\langle b_{\bf l}b^\dagger_{\bf l'}\right\rangle
 =\left[\frac{c}{2}\frac{2}{N}\sum_{\bf k}e^{i{\bf k}({\bf 
 l'}-{\bf l})}\frac{1}{\sqrt{1-r_{\bf k}^2\gamma^2_{\bf k}}}
 \coth\!\left(\! J\beta\varepsilon\sqrt{1-r_{\bf k}^2
 \gamma^2_{\bf k}}\right)\right]^2-
 \frac{c^2}{4}\delta_{\bf ll'},\nonumber\\
&&\label{corr}\\
&&\left\langle {\bf s_ls_m}\right\rangle=
 \left\langle s^z_{\bf l}s^z_{\bf m}\right\rangle\approx
 -\left\langle b_{\bf l}b_{\bf m}\right\rangle
 \left\langle b^\dagger_{\bf l}b^\dagger_{\bf m}\right\rangle
 =-\left[\frac{c}{2}\frac{2}{N}\sum_{\bf k}e^{i{\bf k}({\bf 
 m}-{\bf l})}\frac{r_{\bf k}\gamma_{\bf k}}{\sqrt{1-r_{\bf k}^2
 \gamma^2_{\bf k}}}\coth\!\left(\!J\beta\varepsilon
 \sqrt{1-r_{\bf k}^2\gamma^2_{\bf k}}\right)\right]^2.\nonumber
\end{eqnarray}
\narrowtext\noindent
In the derivation of Eq.~(\ref{corr}) we have taken into account the
constraint of zero site magnetization in the paramagnetic state
\begin{equation}
\left\langle s^z_{\bf l}\right\rangle=\left\langle b^\dagger_{\bf l}
b_{\bf l}\right\rangle-\frac{1}{2}=0,
\label{zsm}\end{equation}
and in calculating $\left \langle b^\dagger_{\bf n}b_{\bf
n}b^\dagger_{\bf n'}b_{\bf n'}\right \rangle$ we have considered only
diagrams corresponding to two independent magnon Green's functions,
neglecting diagrams with interaction wavy and dashed lines between these
two functions.

The index {\bf k} added to the parameter $r$ in Eq.~(\ref{corr}) is
noteworthy. To fulfil constraint (\ref{zsm}) we separate out the terms
with ${\bf k}=0$, the so-called condensation parts, in the sums in spin
correlations (\ref{corr}) and suppose that $r$ in the separated terms
differs from the analogous parameter in terms with ${\bf k}\neq 0$. This
latter parameter is calculated from Eqs.~(\ref{eps}), (\ref{phi}),
(\ref{sums}), and (\ref{omega}), whereas $r_{{\bf k}=0}=r'$ in the
condensation parts is determined from constraint (\ref{zsm}) which can
be written in the form
\begin{eqnarray}
&&\frac{2}{N}\frac{1}{\sqrt{1-r'^2}}\coth\!\left(\! J\beta
 \varepsilon\sqrt{1-r'^2}\right)=\frac{1+c}{c}\nonumber\\
&&\mbox{ }-\frac{2}{N}\sum_{\bf k\neq 0}\frac{1}{
 \sqrt{1-r^2\gamma^2_{\bf k}}}\coth\!\left(\! J\beta\varepsilon
 \sqrt{1-r^2\gamma^2_{\bf k}}\right). \label{zterm}
\end{eqnarray}
In a large lattice and at $T=0$ the condensation parts are equal to
$2m_0$, Eq.~(\ref{pars}), and the sums in Eqs.~(\ref{corr}) for large
$L=|{\bf L}|$ are equal to $2m_0+2^{1/2}\left(\pi L\right)^{-1}$. Thus,
under these conditions
\begin{equation}
\left\langle{\bf s_0s_{L}}\right\rangle\approx (-1)^L\left[m_0+
\left(\sqrt{2}\pi L\right)^{-1}\right]^2,
\label{asymp}\end{equation}
where $(-1)^L=+1$ or $-1$ depending on whether the sites {\bf 0} and
{\bf L} belong to the same or different sublattices. The value of the
sublattice magnetization $m_0\approx 0.3034$ is in good agreement with
the Monte Carlo calculations. \cite{reger,liang} The $1/L$ decay of the
spin correlations to the square of the order parameter at zero
temperature was expected in Ref.~\onlinecite{huse}. For small $L$ the
zero-temperature spin correlations $\left\langle{\bf
s_0s_{L}}\right\rangle$ calculated from Eqs.~(\ref{eps}), (\ref{phi}),
(\ref{sums}), (\ref{corr}), and (\ref{zterm}), are compared with the
results of the projected Monte Carlo method \cite{liang} in
Table~\ref{tabi}\@. The values agree nicely. We notice also that our
spin correlations calculated to the fourth decimal place coincide with
data of the MSWT of Ref.~\onlinecite{sherman}.

For low temperatures the asymptotic behavior of the sums in
Eqs.~(\ref{corr}) for large distances is
$$\frac{1}{\pi^2J\beta\varepsilon}\int\!\!\!\int\frac{d^2k}{k^2+
(2\xi)^{-2}}e^{i\bf kL}\approx\frac{2}{J\beta\varepsilon}\sqrt{\frac{
\xi}{\pi L}}\exp\!\left(-\frac{L}{2\xi}\right),$$
with the correlation length
\begin{eqnarray}
\xi/a&=&\frac{r}{\sqrt{8(1-r^2)}}\nonumber\\
&\approx&\frac{1}{4\sqrt{2}}\left(\frac{m_1}{
 1+2w(1)-4m_0+2m_1}\right)^{1/2}e^{J\beta}.
\label{clength}\end{eqnarray}
Analogous exponential temperature dependences of the correlation length
were obtained earlier in a number of works (see
Refs.~\onlinecite{chakra,takahashi} and references therein). The
exponent $J\beta$ in Eq.~(\ref{clength}) is close to the values
$0.94J\beta$ and $2\pi m_0m_1J\beta\approx 1.1037J\beta$ obtained in
Refs.~\onlinecite{chakra} and \onlinecite{takahashi}, respectively.

In Figs.~\ref{figiv}--\ref{figvi} the energy per spin 
\begin{equation}
E=2J\left\langle{\bf s_0s_a}\right\rangle
\label{energy}\end{equation}
and the static uniform susceptibility
\begin{eqnarray}
\chi&=&\beta\sum_{\bf n}\left\langle s^z_{\bf 0}s^z_{\bf n}\right
 \rangle=\frac{\beta}{3}\sum_{\bf n}\left\langle{\bf s_0s_n}\right
 \rangle\nonumber\\
&=&\frac{c^2\beta}{12}\left[\frac{2}{N}\sum_{\bf k}\coth^2\!\left(J\beta
 \varepsilon\sqrt{1-r^2_{\bf k}\gamma^2_{\bf k}}\right)-1\right]
\label{chi}\end{eqnarray}
calculated with Eqs.~(\ref{freq})--(\ref{phi}), (\ref{sums}),
(\ref{corr}), and (\ref{zterm}) are compared with results of Monte Carlo
simulations, \cite{makivic,okabe} the MSWT of
Refs.~\onlinecite{takahashi,tang} and the exact diagonalization of a
4$\times$4 lattice. \cite{takahashi} The parameters $r$, $\varepsilon$,
and $r'$ obtained in these calculations are given in
Table~\ref{tabii}\@. Respective parameters $\eta$ and $\lambda$ which
define the excitation frequency in the MSWT \cite{takahashi}
$\omega_{\bf k}=\lambda\sqrt{1-\eta^2\gamma^2_{\bf k}}$ [cf.
Eq.~(\ref{omega})] are also given in this table for comparison. The size
dependence of $\chi$ and $E$ calculated with the above formulas is
negligible for large enough lattices --- the difference between values
obtained for a 64$\times$64 lattice and those shown in Figs.~\ref{figiv}
and \ref{figv} for a 20$\times$20 lattice is less than the size of the
symbols in these figures.

As seen from the figures, for temperatures $T\lesssim 0.6J$ the results
obtained with OLA are close to the results of the MSWT
\cite{takahashi,tang} and are in good agreement with the Monte Carlo and
exact-diagonalization data. From Figs.~\ref{figiv} and \ref{figvi} we
see also that energy (\ref{energy}) and the related short-range
correlation are well described by the OLA even for $T \gtrsim J$ where
the MSWT does not work. For the susceptibility in Fig.~\ref{figv} the
agreement is not so good for these temperatures, however the
approximation is able to reproduce the maximum in $\chi$ near $T=0.9J$
and the decrease of the susceptibility for higher temperatures in close
correspondence with experiment and Monte Carlo simulations.

As seen from Table~\ref{tabii}, the parameters $r'$ and $\varepsilon$ of
the OLA are close to the respective parameters $\eta$ and $\lambda/(2J)$
of the MSWT of Ref.~\onlinecite{takahashi} for low temperatures.
Essential deviations begin at $T \approx 0.5J$. We notice that the value
of the parameter $r$ is large at this temperature and grows with growing
$T$. These values of $r$ correspond to a large gap in the excitation
spectrum near ${\bf k}=0$. To understand this result we notice that in
the considered magnetic Brillouin zone the vicinity of the $\Gamma$
point corresponds to two regions in the usual Brillouin zone which is
twice as large as the magnetic one. These regions are located near the
$(0,0)$ and $(\pi/a,\pi/a)$ points. Correspondingly, a branch of
elementary excitations in the usual Brillouin zone transforms into two
branches in the magnetic zone. In Ref.~\onlinecite{sherman} we have
considered how the magnon branch, which is twofold degenerate at $T=0$
in the magnetic Brillouin zone, splits into two branches with growing
temperature. These two branches form a unified branch in the usual
Brillouin zone and are related to the central part and the periphery of
this zone. In Ref.~\onlinecite{sherman} we have indicated that the
branch with the gap near ${\bf k}=0$ in the magnetic Brillouin zone
corresponds to the periphery of the usual zone and the states in this
branch have larger spectral intensities in the magnon Green's function
in comparison with states in the second, gapless branch. In the OLA and
the MSWT of Refs.~\onlinecite{takahashi,tang} the branch does not split
and for high temperatures its behavior is expected to be determined by
states with larger spectral intensities. Thus, the gap mentioned above
is related to the vicinity of the $(\pi/a,\pi/a)$ point in the usual
Brillouin zone.

\section{Summary}
In this paper we used the diagram technique for spin operators
\cite{vaks,izyumov} to calculate Green's functions and observables of
the undoped spin-$\case{1}{2}$ quantum Heisenberg antiferromagnet on a
square lattice. We considered the first corrections -- the one-loop
diagrams -- to the simplest chain diagrams. The obtained equations
resemble the formulas of the modified spin-wave theory of
Refs.\onlinecite{takahashi,tang}. The major difference between these two
approximations is in the calculation of the excitation frequency --- in
the modified spin-wave theory two parameters defining the frequency are
deduced from the constraint of zero site magnetization and a
self-consistency condition, while in the one-loop approximation these
parameters are determined from the values of the diagrams. The one-loop
approximation reproduces the results of the traditional spin-wave
approximation \cite{oguchi,igarashi} on the zero-temperature
renormalization of the excitation frequency, of the
renormalization-group theory \cite{chakra} on the temperature variation
of the correlation length, and of the modified spin-wave theory
\cite{takahashi,tang} on spin correlations. Due to the mentioned
difference in the excitation frequency and some other differences in
formulas, the one-loop approximation is applicable in a wider
temperature range than the modified spin-wave theory. The energy per
spin calculated in this approximation is in good agreement with the
Monte Carlo and small-cluster exact-diagonalization data up to the
temperatures $T\approx 1.2J$, whereas the modified spin-wave
approximation gives quantitatively correct values of this energy in the
range $T\lesssim 0.6J$. In this temperature range both approximations
give values of the static uniform susceptibility in good agreement with
numerical calculations. In the one-loop approximation, the agreement
becomes somewhat worse for larger temperatures. Nevertheless this
approximation is able to describe the maximum in the temperature
dependence of this observable near $T=0.9J$ in close correspondence with
experiment and Monte Carlo simulations, while in the modified spin-wave
theory \cite{takahashi,tang} the susceptibility grows monotonously with
temperature.

\acknowledgements
This work was partially supported by the ESF grant No.~4022 and by the
WTZ grant (Project EST-003-98) of the BMBF.

\begin{figure}\caption{Elements of diagrams. See text for an explanation
of the different lines.}
\label{figi}\end{figure}

\begin{figure}\caption{Chain diagrams for $D_{\bf ll'}(\tau)$.}
\label{figii}\end{figure}

\begin{figure}\caption{The lowest-order corrections to self-energies
and interaction lines.}\label{figiii}\end{figure}

\begin{figure}\caption{The energy per spin obtained in the Monte Carlo
simulation \protect\cite{makivic} ($\bullet$), in the modified spin-wave
theory of Refs.~\protect\onlinecite{takahashi,tang} ($\circ$) and in the
one-loop approximation of the present work ($\times$). In the modified 
spin-wave and one-loop calculations a 20$\times$20 lattice was used.}
\label{figiv}\end{figure}

\begin{figure}\caption{The static uniform susceptibility obtained in the
Monte Carlo simulation for a 12$\times$12 lattice \protect\cite{okabe}
($\bullet$), in the modified spin-wave theory of
Refs.~\protect\onlinecite{takahashi,tang} ($\circ$) and in the one-loop
approximation of the present work ($\times$). In the modified spin-wave 
and one-loop calculations a 20$\times$20 lattice was used.}
\label{figv}\end{figure}

\begin{figure}\caption{The energy per spin obtained by exact
diagonalization of a 4$\times$4 lattice \protect\cite{takahashi}
($\bullet$) and in the one-loop approximation ($\times$).}
\label{figvi}\end{figure}

\begin{table}
\caption{The zero-temperature spin correlations $\left\langle{\bf s_0
s_L}\right\rangle$ obtained 
with the one-loop approximation (OLA) for a 20$\times$20
lattice in comparison with the projected Monte Carlo data (PMC).
\protect\cite{liang}
}
\label{tabi}
\begin{tabular}{|c|r|r|}
{\bf L}   & PMC     & OLA     \\ \hline
$(a,0)$   & -0.3348 & -0.3354 \\
$(a,a)$   &  0.2028 &  0.2016 \\
$(2a,0)$  &  0.1772 &  0.1751 \\
$(2a,a)$  & -0.1671 & -0.1648 \\
$(2a,2a)$ &  0.1475 &  0.1454 \\
$(3a,0)$  & -0.1491 & -0.1461 \\
$(3a,a)$  &  0.1430 &  0.1404 \\
\end{tabular}
\end{table}

\begin{table}
\caption{Parameters $r$, $\varepsilon$, $r'$ in the one-loop
approximation (OLA) and the respective parameters of the modified
spin-wave theory \protect\cite{takahashi,tang} (MSWT). In both 
calculations a 20$\times$20 lattice was used.
}
\label{tabii}
\begin{tabular}{|l||r|r|r||r|r|}
  &\multicolumn{3}{c||}{OLA}&\multicolumn{2}{c|}{MSWT}\\ 
  \hline
$T/J$ &    $1-r$     &   $\varepsilon$ &  $1-r'$
  &   $1-\eta$  &   $\lambda/(2J)$\\   \hline
0.01 & 0.8634E-04 & 1.1582 & 0.4192E-04 & 0.4193E-04 & 1.1582 \\
0.1  & 0.8633E-03 & 1.1583 & 0.3188E-03 & 0.3197E-03 & 1.1583 \\
0.2  & 0.1860E-02 & 1.1561 & 0.6434E-03 & 0.6480E-03 & 1.1570 \\
0.3  & 0.6692E-02 & 1.1490 & 0.9904E-03 & 0.1038E-02 & 1.1516 \\
0.4  & 0.2528E-01 & 1.1385 & 0.1276E-02 & 0.1566E-02 & 1.1374 \\
0.5  & 0.5953E-01 & 1.1257 & 0.1508E-02 & 0.2404E-02 & 1.1097 \\
0.6  & 0.1033     & 1.1101 & 0.1735E-02 & 0.4081E-02 & 1.0629 \\
0.7  & 0.1511     & 1.0928 & 0.1976E-02 & 0.8882E-02 & 0.98897 \\
0.8  & 0.1995     & 1.0752 & 0.2230E-02 & 0.2930E-01 & 0.87371 \\
0.9  & 0.2467     & 1.0589 & 0.2498E-02 &            & \\
1.0  & 0.2918     & 1.0446 & 0.2774E-02 &            & \\
1.1  & 0.3344     & 1.0327 & 0.3055E-02 &            & \\
1.2  & 0.3743     & 1.0234 & 0.3336E-02 &            & \\
\end{tabular}
\end{table}

\end{document}